\documentclass[aps,11pt,prx,longbibliography]{revtex4-1}
\usepackage[utf8]{inputenc}
\usepackage[T1]{fontenc}
\usepackage{lmodern}
\usepackage[francais, english]{babel}
\usepackage{amssymb}
\usepackage{amsmath,amsfonts,color,bm}
\usepackage{esvect}
\usepackage{graphicx}
\usepackage{fancyhdr}
\usepackage{hyperref}
\usepackage{xr-hyper}

\newcommand{\Pe}[0]{{\rm Pe}}

\usepackage{pst-tree}
\usepackage{epstopdf}
\usepackage{listings}
\usepackage{vmargin}  
\usepackage{latexsym}
\usepackage{graphicx}
\usepackage{amssymb}
\usepackage{epstopdf}
\usepackage{wrapfig}

\begin{document}
\title{Mechanically activated ionic transport across 2nm carbon nanotubes}
\title{Mechanically activated ionic transport across single digit carbon nanotubes}

\author{Alice Marcotte, Timoth\'ee Mouterde, Antoine Nigu\`es, Alessandro Siria$^{\star}$, Lyd\'eric Bocquet$^{\star}$}
\affiliation{Laboratoire de Physique de l'Ecole normale Sup\'erieure, ENS, Universit\'e PSL, CNRS, \\
Sorbonne Universit\'e, Universit\'e de Paris, 75005 Paris, France}
\email{lyderic.bocquet@ens.fr, alessandro.siria@ens.fr}

\begin{abstract}

{\bf Fluid and ionic transport at nanoscale recently highlighted a wealth of exotic behaviours \cite{Bocquet2010,SDN2019,Bocquet2020,Majumder2005,Noy2006,
Noy2016,Secchi2016,Noy2017,Siria2013,Siria2017,Radenovic2016,  Radenovic2016_2,Keyser2017,Karnik2013}
.  However, the artificial nanofluidic devices \cite{Esfandiar2017,Radha2016,Radha2018,Mouterde2019} are still far from the advanced functionalities existing in biological systems, such as electrically and mechanically activated transport \cite{Anishkin2014,Cox2019}. Here we focus on the ionic transport through 2 nm-radius individual multiwalled carbon nanotubes (CNT), under the combination  of mechanical and electrical forcings. Our findings evidence mechanically activated ionic transport under the form of an ionic conductance which depends quadratically on the applied pressure. Our theoretical study relates this behaviour with the complex interplay between electrical and mechanical drivings, and shows that the superlubricity of CNT \cite{Lindsay2011,Noy2017,Noy2016,Noy2006,Majumder2005,Secchi2016} is a prerequisite to attain mechanically activated transport. The pressure sensitivity shares similarities with the response of biological mechanosensitive ion channels \cite{Anishkin2014,Cox2019} observed here for the first time in an artificial system. This paves the way to build new active nanofluidic functionalities inspired by the complex biological machinery. 
}

\end{abstract}

\maketitle

Nanofluidics, the field studying transport of fluids at the nanoscales, has now reached a stage of maturity. The manifold systems which have been developped over the recent years, from nanopores, nanotubes, to angstr\"om scale channels, together with the new instrumentation developped, now allow investigating fluid transport at the frontier of hydrodynamics. This points to key knowledge gaps in the so-called single digit nanopores with size below ten nanometers \cite{SDN2019}. 

Still, such artificial systems remain far from the impressive complexity of the biological ionic machinery, which exhibits a plethora of advanced transport functionalities. In particular, the recent identification of mechanotransduction channels 
has highlighted complex response involving excitatory mechanically activated ionic current under mechanical pressure or stretching \cite{Anishkin2014,Cox2019}. Reproducing such advanced response in artificial systems was up to now unreachable.



\begin{figure*}[h!]
\includegraphics[scale=1]{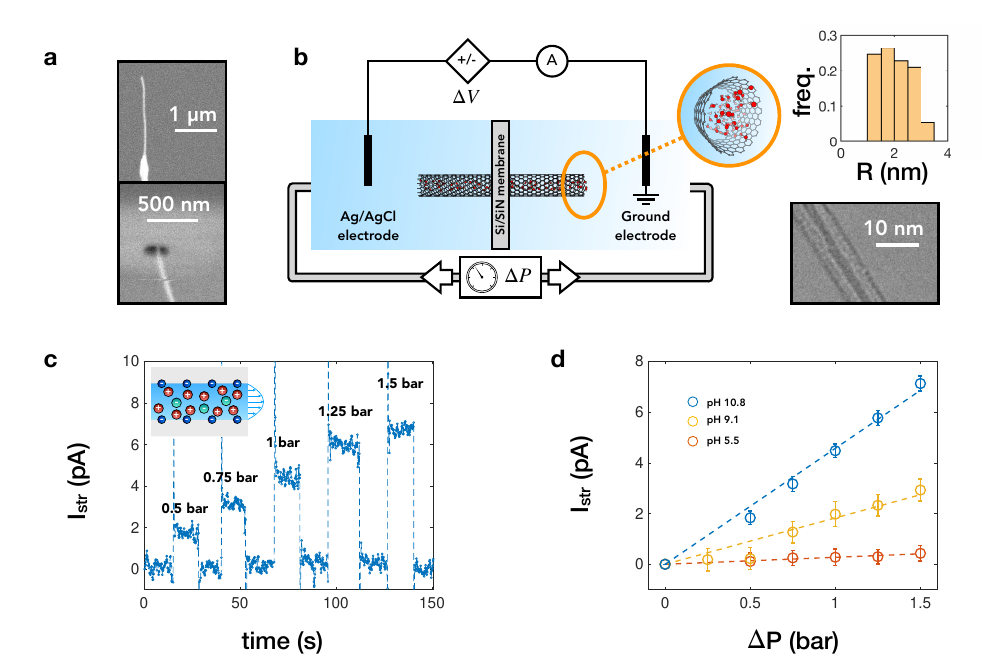} 
\caption{\label{setup} \textbf{Experimental setup and pressure-driven current without bias voltage ($\Delta V = 0$).} \textbf{a,} Top: CNT glued at the apex of an electro-chemically etched tungsten tip (SEM imaging). Bottom: CNT introduced into the hole drilled in a Si/SiN membrane (SEM imaging) \textbf{b,} Left: schematic of the nanofluidic cell for pressure and voltage-driven current measurements through a single CNT. Right, top: inner radii distribution for a sample consisting in 57 CNTs observed under TEM imaging. Right, bottom: a CNT observed under TEM imaging.  \textbf{c,} Ionic current as a function of time, at a salt concentration of 1 M and pH 10.8 without bias voltage ($\Delta V = 0$). A sliding average of the current over four data points is performed. Current overshoots once the pressure is applied, and then saturates to a steady-state value. Inset: Sketch for the pressure-driven current through a CNT. \textbf{d,}~Streaming current as a function of the pressure drop $\Delta P$ at different pHs without applied voltage ($\Delta V = 0$). For each pH, the dashed line represents the best linear fit. The error bars in \textbf{d} represent the standard error over five experimental replicates.}
\end{figure*}

In the quest to reproduce such advanced response, we investigate here the non-linear transport in individual 'single-digit' CNT with radius  $\approx$ 2 nm and explore how mechanical pressure does affect the ionic transport in these channels. CNT are unique materials in that they highlight ultra-low water friction on their surface \cite{Secchi2016} and electro-dynamic couplings are accordingly expected to be exalted. However, probing transport in individual CNT with nanometric radius under coupled voltage and pressure drops has been so far a considerable challenge, notably in terms of fabrication and sensitivity, and most experimental studies in such tiny NT have been restricted up to now to electric forcings \cite{Lindsay2010,Lindsay2011,Strano2013,Jourdain2017}. 
To overcome this bottleneck, we build on the transmembrane nanotube methodology, initially developed for the measurement of the ionic transport  with larger 10 - 50 nm radii nanotubes, \cite{Siria2013}, but here applied to 2 nm nanotubes. 
The experimental set-up is shown in Fig. \ref{setup}b: an individual arc-discharge CNT (SIGMA ALDRICH,  Product number: 406074) connects two macroscopic fluid reservoirs through a 50 nm thick silicon-nitride (SiN) membrane (NORCADA). We investigated CNTs with inner radii of $R\approx 2.0 \pm 0.6 $ nm  and lengths of $ L \approx $ 1 $\mu$m. The length of the CNTs is measured during the fabrication procedure under Scanning Electron Microscope (SEM) imaging (see Fig. \ref{setup}a) and the inner radii distribution for the CNT has been obtained by Transmission Electron Microscopy (TEM) observations and is displayed on Fig. \ref{setup}b. Transmembrane individual CNT (t-CNT) devices are realized by \textit{in situ} SEM nanomanipulation, using the method previously described in \cite{Siria2013} (see Methods for more details). The membrane is then sealed between two reservoirs, which are filled with a potassium chloride (KCl) solution at various concentrations (denoted as $c$) and pHs. One Ag/AgCl electrode is immersed in each reservoir, and connected to a patch-clamp amplifier (Axopatch 200B) which allows for ionic current measurements with a current resolution of 0.1 pA. Both reservoirs can be connected to a pressure controler, enabling us to impose a pressure drop ($\pm$ 0 - 2 bar) in both directions across the membrane. This setup allows measuring the ionic current driven by a pressure drop $\Delta P$ and the combined effect of additional forcing  $\Delta V$  applied along the t-CNT. In this study, three independent t-CNT devices with similar geometries were studied, and gave the same response as described below. Their IV characteristics at c =  1 M and pH 5.5 are given by Supplementary Figure 1.

In a typical experiment, we measured the pressure-driven component of the ionic current $I$, referred to as the streaming current and defined here as $ I_{\rm str} = I(\Delta P, \Delta V) - I(\Delta P = 0, \Delta V)$, see Methods. 
This current results from the transport of (excess) ions carried by a pressure-induced flow (inset Fig. 1c) and probes the interplay between ion and mass transport. A pressure drop between 50 mbar and 1.5 bar is applied in both directions across the t-CNT. We do not observe a temporal evolution of the system and the conductance is the same before and after pressure measurements are performed. Let us focus first on the pressure-response of the system in the absence of voltage bias ($\Delta V=0$). As shown on Figure \ref{setup}c and on Supplementary Figure 2 
for $c$ = 1 M and $\Delta V = 0$, once the pressure drop is applied, the streaming current overshoots and then quickly relaxes to a stationary value. 

We report in the Figure \ref{setup}d, the stationary value reached by $I_{str}$ as a function of the applied pressure $\Delta P$. As usually expected for such measurements without applied voltage, the streaming current varies linearly with the applied pressure, $I_{\rm str} \propto \Delta P$, see Fig. \ref{setup}d, and the slope increases with pH. The sign of the current is furthermore consistent with an excess transport of positive charges flushed by water flow. This response is in agreement with results obtained for  CNT with larger diameters \cite{SecchiPRL} and was explained in terms of hydroxide adsorption on the CNT walls. Another contribution can originate from the differential mobility of the ions in confinement, as discussed recently \cite{Mouterde2019}.

\begin{figure*}[h!]
\includegraphics[scale=1]{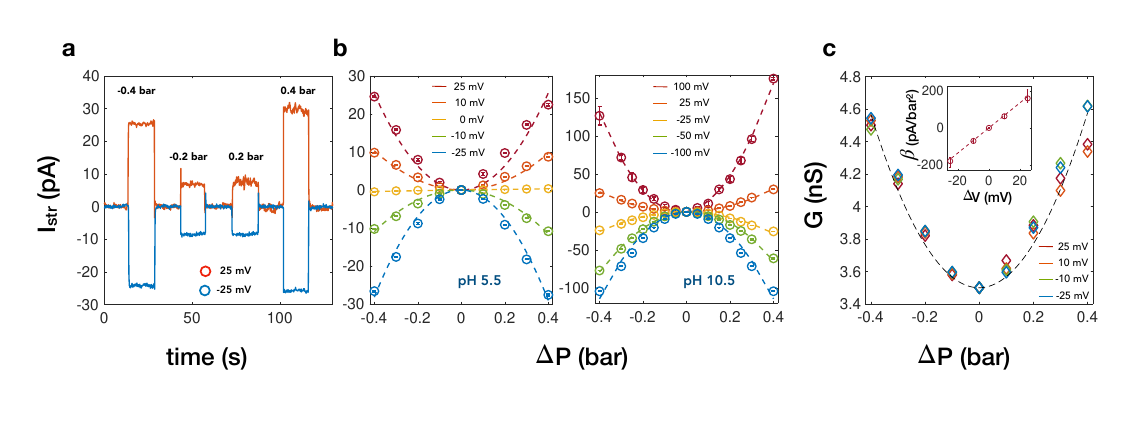} 
\caption{\label{non_linear} \textbf{Pressure-driven current while applying a bias (experiments): a,} Streaming current as a function of time, at $c$ = 1 M and at pH 10.5 while applying a bias voltage $\Delta V$ = 25 mV (red curve) and $\Delta V$ = -25 mV (blue curve). \textbf{b,} Streaming current as a function of the pressure drop $\Delta P$ while applying a bias voltage $\Delta V \in \left[-100\text{ mV},100\text{ mV}\right]$, at $c$ = 1 M and at pH = 5.5 (left) and pH = 10.5 (right). For each applied bias, the dashed line represents the best parabolic fit. \textbf{c,} Mechano-sensitive conductance  $G$ as a function of the pressure drop $\Delta P$  for different bias voltages $\Delta V \in \left[-25\text{ mV},25 \text{ mV}\right]$, at $c$ = 1 M and pH = 5.5. The dashed line represents the best parabolic fit. Inset: Second-order coefficient from the best parabolic fit in \textbf{b}, as a function of the bias $\Delta V$, for $c$ = 1 M and pH = 5.5. The dashed line represents the best linear fit.  Error bars represent: \textbf{b}, standard error over 25 replicates; \textbf{c}, inset: uncertainty in the fit value obtained in \textbf{b}.
}
\end{figure*}

In the present study, we rather focus on the pressure response in the presence of a constant bias voltage $\Delta V$, typically $\Delta V\in [-100$ mV$;100$ mV$]$. Figure \ref{non_linear}a shows the time-dependent response of the streaming current while applying pressure drops of $\pm$200 mbar and $\pm$400 mbar, when a bias voltage of 25 mV and of -25 mV is imposed along the CNT. We measure that for a given bias voltage, the streaming current is of the same sign independently of the sign of the pressure drop. This is best illustrated in Fig. \ref{non_linear}b, which shows that the 
streaming is now a non-linear function of the pressure drop. This response is evidenced for all investigated systems, salt concentrations and pHs, see Supplementary Figures 3, 4 and 5. 
The experimental data is very well described by a second order polynomial function of the pressure drop:

\begin{equation}
I_{\rm{str}}(\Delta P)= \mu_{EO} \Delta P + \beta (\Delta V)\times \Delta P^2,
\end{equation} 
where the first term is the (linear) electro-osmotic term with mobility $\mu_{EO}$ and the second quadratic term is the dominant contribution to the current, Fig. \ref{non_linear}b. Furthermore, the coefficient $\beta (\Delta V)$ is shown to depend linearly on the gating voltage, $\Delta V$, as indicated on the inset of Fig. \ref{non_linear}c. 

Gathering the streaming and bare conductance contributions to the ionic current, one finds that the total current can be rewritten
as 
\begin{equation}
 I_{\Delta P}(\Delta V) = \mu_{EO} \times \Delta P  + G(\Delta P)\times  \Delta V
 \label{IDPDV}
 \end{equation}

 where the conductance is now a quadratic function of the applied pressure, $G(\Delta P)=G_0+G_2 \times \Delta P^2$. This result highlights that narrow carbon nanotubes display mechano-sensitive ionic conduction, with a conductance which is tuned by the mechanical pressure. Figure \ref{non_linear}c represents the conductance $G(\Delta P)$ as a function of the pressure drop $\Delta P$, here for a solution of salt concentration of 1 M and pH 5.5. The collapse of the $G$ curves for various $\Delta V$  shows that the pressure-dependent conductance is the physically relevant quantity here.  This behaviour was measured for all investigated pHs and salt concentrations. Supplementary Figure 6 shows that $G_2$ increases with salt concentration (for both pH 5.5 and pH 10.5) with only a weak dependence on pH. This  suggests that this result does not stem from a surface charge effect contrary to the case of the -- much smaller -- streaming currents without applied bias displayed in the Figure \ref{setup}c. In the latter case, the pH-dependent surface charge of the CNT is expected to be the main contribution to the bare streaming current\black{}. Furthermore, both the effects of the stretching of the supporting membrane, and the change of radius of the CNT can be discarded as the origin of this phenomenon, since both are negligible due to the high stiffness of the materials (in contrast to explanations for the mechano-sensitive response of biological channels \cite{Anishkin2014, Cox2019}). 
Interestingly the observed response in CNT is also very different from that observed recently in 2D Angstr\"om slits with carbon confinement \cite{Mouterde2019}, where only a linear dependence of the ionic current as a function of pressure drop could be observed whatever the applied voltage. Here the response reveals a robust, quadratic, pressure-dependence of the  conductance. Interestingly, 
such pressure-sensitive behaviour shares similarities with the mechanosensitive response of biological channels 
 \cite{Anishkin2014, Cox2019}. 
In the latter systems, a pressure drop converts into a ionic current, showing
a non-linear dependence on the applied pressure.  

In order to rationalize this behaviour, we have developed a theoretical model capturing the main ingredients of ionic transport in narrow carbon nanotubes. 
A key feature is the low friction expected for water on the CNT surface, which results in fast flows inside the CNT under imposed pressure drop. We anticipate that this is at the root of the strong interplay between the ionic transport and mechanical forcings.

The theoretical description proceeds with a modified Poisson-Nernst-Planck-Stokes (PNPS) description for the ion transport inside the CNT, building on the framework introduced in \cite{Mouterde2019}; the details of the derivation are provided in the Supplementary Note 1. In short, we consider the 1-dimensional transport of ions inside the CNT, under the combined effects of the local electric field and (fast) flow of water. Ion mobilities are not necessarily equal and we account for the friction of ions on both water and surface walls. Last but not least, water slippage on the CNT surface is assumed. The PNPS framework writes the conservation equations for each specie and water along the channel.
 
Under limited approximations, this extended PNPS model can be solved analytically in 1D and one obtains an analytical prediction for the ionic current under the combined effects of pressure and voltage drops. Remarkably the theory predicts that the conductance is a function of a P\'eclet number $\Pe$, 
defined as
\begin{equation}
\Pe=\frac{Q_w}{D}\times \frac{L}{\pi R_0^2}
\end{equation}
 where $Q_w$ is the water flow rate and $D$ a typical ion diffusion coefficient. Note that the definition of the P\'eclet number involves the length $L$ (and not the radius $R$) of the channel. To linear order in $\Delta V$ as in the experiments, but keeping the full dependence on the P\'eclet number, one predicts 
 
 \begin{equation}
I =   \mu_{EO} \times \Delta P+G(\Pe) \times \Delta V+\dots
\label{IePe1}
\end{equation}
where $ \mu_{EO}$ is an electro-osmotic mobility, which is a function of the various kinetic parameters -- see Supplementary Information for details --, and $G(\Pe)$ is a P\'eclet-dependent conductance. The theory predicts a  transparent expression for the conductance as:
\begin{equation}
G(\Pe)=G_0 \times {{\Pe\over 2}\over \tanh {\Pe\over 2}}
\label{GPe}
\end{equation}
which can be expanded as $G(\Pe)=G_0\times (1+ {\Pe^2\over 12} + \dots)$. 
One expects the pressure-driven contribution to the flow to be the main contribution to the P\'eclet number, so that $\Pe= {Q_w L/D\pi R_0^2}={K_{\rm app}\over D} \times \Delta P$, with $K_{\rm app}$ the apparent permeance of the CNT. The ionic conductance $G(\Pe)$ is now a  function of the pressure drop, varying quadratically for low $\Delta P$. This is in full agreement with experimental measurements, see Fig.\ref{non_linear} and Eq. (\ref{IDPDV}). 
Physically, this square-like dependence of the conductance on the P\'eclet number results from the balance between electric and mechanical forces. This leads to an overall increase of the ionic concentration in the nanotube for both $\Delta P>0$ and $\Delta P<0$ -- as can be verified from the direct calculation of the ionic density profiles in the channel, see Supplementary Note 1. 
--, hence an increase in the conductance of the nanotube which is symmetric for positive and negative pressure drops. 
A key remark is that the P\'eclet number (hence flow) cannot be neglected here to describe ionic transport in tiny CNTs, in strong contrast to usual channels: typical values for the P\'eclet number may be inferred from the current-pressure characteristics such as in Fig. \ref{non_linear}b, yielding $\Pe \sim 1-10$ depending on the conditions and nanotube size. In contrast, a Poiseuille flow prediction with no-slip boundary condition on the surface would yield a P\'eclet number in the range $\Pe\sim 10^{-2}$ (considering a 2 nm-radius channel under a maximum pressure drop of 0.5 bar as applied here). But this estimate for $\Pe$ is amplified by a factor $1+8b/R$ when surface slippage, with slip length $b$, occurs on the CNT surface. Hence, large slippage, {\it i.e.} low surface friction on the CNT surface, leads to a strong sensitivity of the conductance on the pressure drop, as highlighted in 
Eq.(\ref{GPe}). This makes the slippery carbon nanotube unique to obtain the mechanosensitive conductance.

In order to further assess these predictions, we solved numerically the full PNPS ionic transport model using a finite-element approach (COMSOL) (see Supplementary Note 2). Overall, these numerical calculations fully confirm the results of the analytical model described above. 
A linear dependence of the streaming current on the pressure drop is recovered when no bias voltage is applied, and the magnitude of the streaming current is as expected much larger in the low friction regime than in the high friction regime, see Supplementary Figure 7.

\begin{figure*}[h!]
\includegraphics[scale=1]{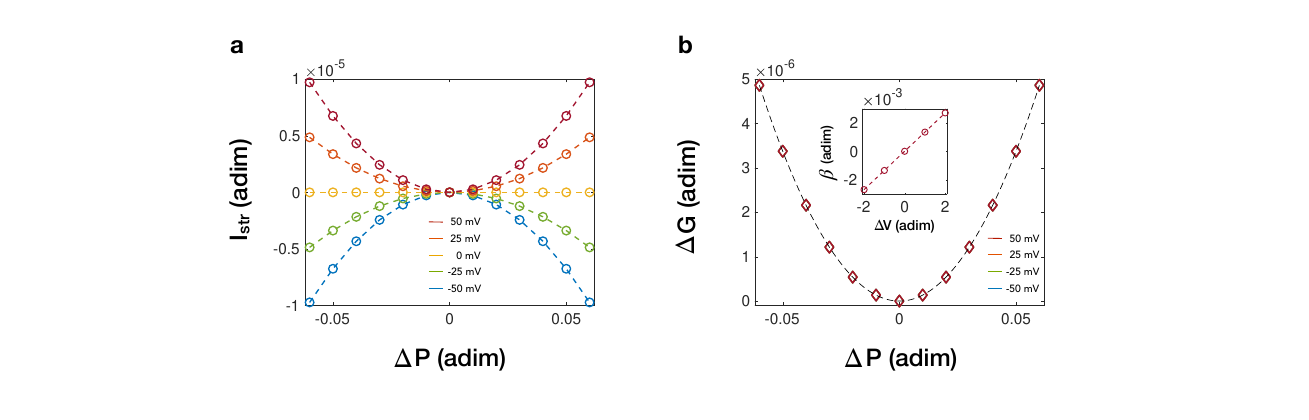} 
\caption{\label{simus} \textbf{Theoretical analysis of the electro-hydrodynamic coupling.} \textbf{a,} Streaming current as a function of the pressure drop $\Delta P$ with different bias voltages $\Delta V\in \left[-50\text{ mV},50 \text{ mV}\right]$. For each bias, the dashed line represents the best parabolic fit (numerical simulations). \textbf{b,} Mechano-sensitive conductance $\Delta G=G(\Delta P)-G_0$ as a function of the pressure drop $\Delta P$, for different bias voltages  $\Delta V \in \left[-50\text{ mV},50 \text{ mV}\right]$. The dashed line represents the best parabolic fit.  Inset: second-order coefficient from the best parabolic fit in \textbf{a}, as a function of the bias $\Delta V$. The dashed line represents the best linear fit. }
\end{figure*}

Now, applying simultaneously voltage and pressure drops,
the numerical calculations highlight a quadratic response of the conductance as a function of the pressure drop, Fig.~\ref{simus}, and furthermore show
that this response is intimately connected to a low surface friction ({\it i.e.} strong slippage); see the high versus low friction response in Supplementary Figure 8, and the $G_2$ dependency on the slip length $b$ on Supplementary Figure 9.
A linear response in $\Delta P$ is recovered in the high friction regime, and is reminiscent of the results reported in \cite{Mouterde2019} in
Angstr\"om slits (slippage was shown to be larger in nanotubes than on flat graphite \cite{Secchi2016}). But a quadratic response in $\Delta P$ is obtained for the low friction surface as observed here for the CNT.

Overall we unveil that the conductance of the CNT is quadratically modulated by the applied pressure. It is interesting to remark that this result is reminiscent of the response of mechanosensitive biological channels.
However, in the case of mechano-sensitive biological channels the mechanism at stake rather involves the sensing of changes in membrane tension coupled to conformational changes in the channel protein structures. As discussed above, such mechanism does not occur for the transmembrane CNT considered here. 
Our results rather point to the unique properties of the 'single digit' CNTs to obtain such {pressure-sensitive} response, whereby a pressure-dependent  accumulation of ionic species occurs in the CNT under a voltage bias.  The low friction of water on the carbon surface is accordingly shown to be a prerequisite to the quadratic pressure dependence of the ionic response in the CNT. Single digit CNTs are thus able to respond to very local pressure signal, with nanometric spatial resolution.

Hence, this artificial system reproduces a pressure-sensitive response. 
 Since the device properties are robust and well calibrated, it provides a unique platform to study such advanced response in an artificial system. In addition, the possibility of inserting CNTs into lipid membranes, as developed in Refs \cite{Noy2017,Wu2013}, provides promising applications to conceive new types of ultra-local pressure sensors using {\it e.g.} assemblies of vesicles as electric compartments with pressure-modulated potential. This would constitute a promising building-block to fabricate sensitive soft materials, enabling one to reproduce artificial sensing devices for (light) touch or proprioception. Such hybrid devices offer manifold perspectives in the development of iontronic systems for which the response intensity can be tuned. Finally, our discovery should benefit to the flourishing field of bottom-up artificial cells fabrication for synthetic ion channels design, which mechanosensitive response can be rationally engineered, via our theoretical model, by tuning the nanochannel nature, length or size.

\noindent{\bf Acknowledgements}\\
A.S. acknowledges
funding from the EU H2020 Framework
Programme/ERC Starting Grant agreement number 637748-NanoSOFT.
L.B. acknowledges funding from the EU H2020 Framework Programme/ERC
Advanced Grant agreement number 785911-Shadoks and 
ANR project Neptune. 
L.B. and A.S. acknowledge support from the Horizon 2020 program through Grant No. 766972-
FET-OPEN-NANOPHLOW.

\noindent{\bf Author contributions} \\
L.B. and A.S. designed and directed the project.
A.M., A.S., and A.N. fabricated the devices. A.M. performed the measurements with
inputs from the other authors. All authors analyzed the data and contributed to discussions. 
L.B. performed the theoretical analysis and A.M. the numerical modelling. 
A.M., A.S and L.B. wrote the manuscript with inputs from T.M. 

\noindent{\bf Data availability}\\
The data that support the plots within this paper and other findings of this study are provided with this paper as source data and in Supplementary Information. 

\noindent{\bf Competing interests}\\
The authors declare no competing interests.

\newpage
\section*{Figures}

\newpage
\newpage

\section*{Methods}

\textbf{Device fabrication.} Our devices were fabricated following previously reported procedures \cite{Siria2013,SecchiPRL}. Briefly, a single multiwall carbon nanotube (SIGMA ALDRICH, Product number: 406074) glued at the apex of an electrochemically etched tungsten tip is inserted into a 200 nm hole drilled in a 50 nm thick silicon nitride (SiN) membrane hanging on a centimetric silicium (Si) frame of 500 $\mu$m in thickness (NORCADA). The insertion is performed in the chamber of a Scanning Electron Microscope using a built-in nano manipulator consisting in three axis piezo-inertial step motors of nanometric precision controlled by an outside joystick. The sealing of the tube in the membrane nanohole is ensured by localized cracked naphthalene deposition. Eventually, the sealed nanotube is cut free from the tungsten tip by highly energetic electron beam exposure in water vapor. The centimetric Si/SiN chip is then squeezed between two macroscopic reservoirs filled with a potassium chloride (KCl) solution of various concentrations. All steps of this procedure follow those described in detail in the Supplemental Materials of \cite{Siria2013}.

\textbf{Streaming current measurements.} Ag/AgCl electrodes are immersed into the reservoirs and connected to an external patch-clamp amplifier (Axopatch 200B, Molecular Devices) which allows for electrical forcing  as well as for electrical measurements, with a resolution in the tenths of picoamperes. The ionic current is measured versus time, while steps of pressure ranging from 50 mbar to 1.5 bar are applied via a pressure pump (AF1, Elveflow). Each step typically lasts 15 s, and the delay between two successive steps is 15 s. The streaming current is defined as the difference between the plateau value reached by the ionic current while the pressure is applied, and its value when the pressure is released. For each drop of pressure, this measurement is performed between 2 and 25 times. Supplementary Figure 2 
gives an example of an ionic current measurement as a function of time, when no bias voltage is applied.

\end{document}